\documentclass[letterpaper]{article} 

\usepackage{aaai2026}  
\usepackage{times}  
\usepackage{helvet}  
\usepackage{courier}  
\usepackage[hyphens]{url}  
\usepackage{graphicx} 
\usepackage{color}
\usepackage{url}
\usepackage{framed}
\usepackage{adjustbox}
\usepackage[most]{tcolorbox}
\usepackage{lipsum}
\usepackage{multicol}
\usepackage{changepage}
\usepackage{enumitem}

\usepackage{multicol}
\usepackage{multirow}
\usepackage{amsmath}

\usepackage{natbib}  
\usepackage{caption} 
\usepackage{algorithm}
\usepackage{algorithmic}

\usepackage{newfloat}
\usepackage{listings}
\DeclareCaptionStyle{ruled}{labelfont=normalfont,labelsep=colon,strut=off} 
\floatstyle{ruled}
\newfloat{listing}{tb}{lst}{}
\floatname{listing}{Listing}
\title{ScholarSearch: Benchmarking Scholar Searching Ability of LLMs}
\author{
    Junting Zhou\textsuperscript{\equalcontrib},
    Wang Li\textsuperscript{\equalcontrib },
    Yiyan Liao\textsuperscript{\equalcontrib }, \\
    Nengyuan Zhang,
    Tingjia Miao,
    Zhihui Qi,
    Yuhan Wu\textsuperscript{\dag},
    Tong Yang\thanks{Corresponding author; Contact ways:
    yangtong@pku.edu.cn; \\ yuhan.wu@pku.edu.cn}
}
\affiliations{
    Peking University, Beijing, China\\

}

\usepackage{bibentry}

\begin{document}

\maketitle

\begin{abstract}
Large Language Models (LLMs)' search capabilities have garnered significant attention.  Existing benchmarks, such as OpenAI's BrowseComp, primarily focus on general search scenarios and fail to adequately address the specific demands of academic search. These demands include deeper literature tracing and organization, professional support for academic databases, the ability to navigate long-tail academic knowledge, and ensuring academic rigor. 
Here, we proposed \textbf{ScholarSearch}, the first dataset specifically designed to evaluate the complex information retrieval capabilities of Large Language Models (LLMs) in academic research.  ScholarSearch possesses the following key characteristics: \textbf{Academic Practicality}, where question content closely mirrors real academic learning and research environments, avoiding deliberately misleading models; \textbf{High Difficulty}, with answers that are challenging for single models (e.g., Grok DeepSearch or Gemini Deep Research) to provide directly, often requiring at least three deep searches to derive; \textbf{Concise Evaluation}, where limiting conditions ensure answers are as unique as possible, accompanied by clear sources and brief solution explanations, greatly facilitating subsequent audit and verification, surpassing the current lack of analyzed search datasets both domestically and internationally; and Broad Coverage, as the dataset spans at least 15 different academic disciplines. Through ScholarSearch, we expect to more precisely measure and promote the performance improvement of LLMs in complex academic information retrieval tasks. The data is available at: \url{https://huggingface.co/datasets/PKU-DS-LAB/ScholarSearch}
\end{abstract}

    

\begin{figure*}[htbp]
\centering
\includegraphics[width=1\textwidth]{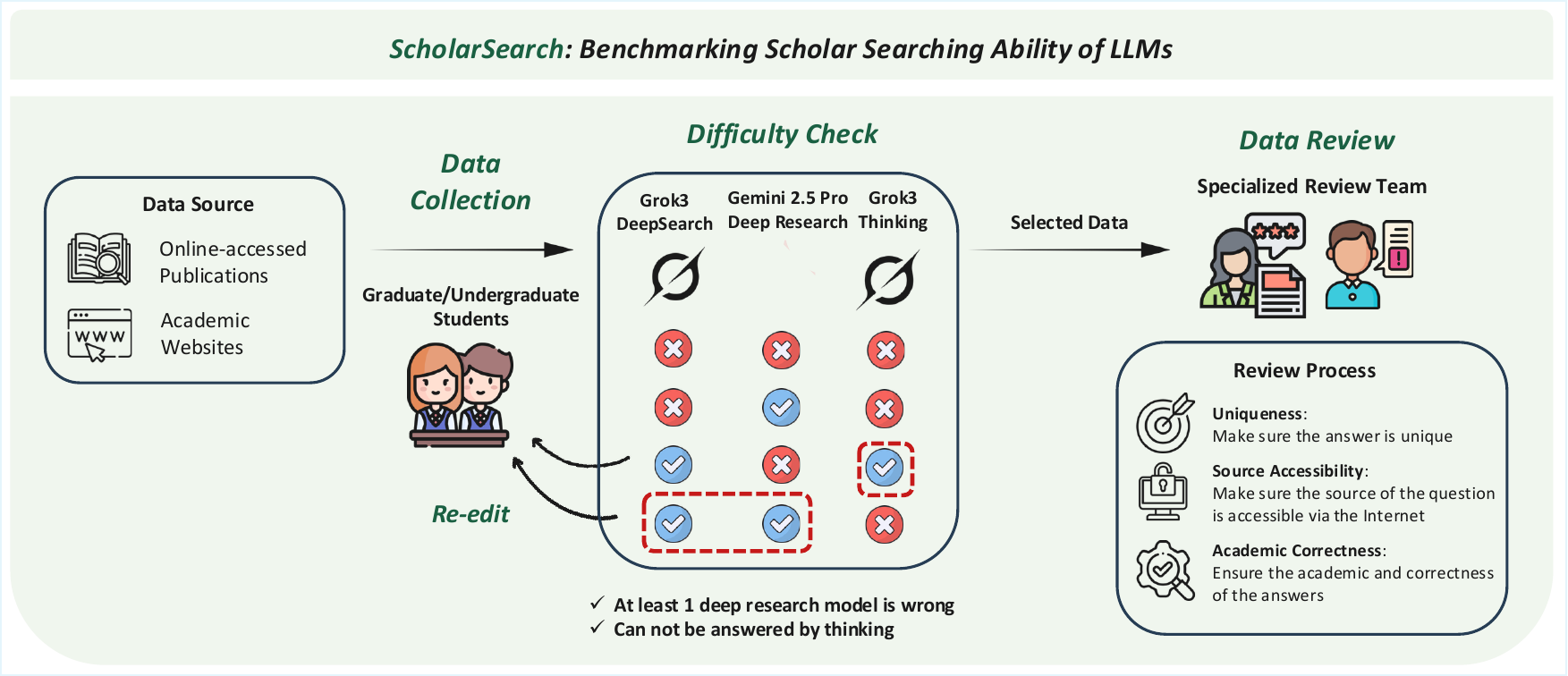} 
\caption{The data collection pipeline of ScholarSearch.  Data sourced by students from academic publications or websites is first filtered using several LLMs to select for difficult questions. A specialized team then reviews this selected data for uniqueness, source accessibility, and academic correctness.}
\label{main}
\end{figure*}

\section{Introduction}
Large Language Models (LLMs) have demonstrated impressive capabilities in various natural language processing tasks, including reasoning, generation, and summarization. While advanced reasoning models possess a considerable repository of accumulated knowledge, a critical limitation persists: their inability to acquire and integrate the latest information in real-time. In the dynamic landscape of scientific research, the capacity for large language models to accurately extract critical, up-to-date information from vast online data sources is paramount. This real-time information retrieval is indispensable for staying abreast of new discoveries, validating hypotheses, and pushing the boundaries of knowledge. Without robust search and retrieval mechanisms, even the most sophisticated LLMs risk becoming quickly outdated in fast-evolving academic fields.

Current efforts to evaluate LLM capabilities in information retrieval primarily fall into two categories: Academic Benchmarks and Browsing Benchmarks. Academic benchmarks, such as GPQA\cite{zhang2023gpqa}, SuperGPQA\cite{pteam2025supergpqascalingllmevaluation}, MMLU\cite{hendrycks2021mmlu}, and MMLU-pro\cite{wang2024mmluprorobustchallengingmultitask}, primarily assess an LLM's understanding and reasoning over pre-existing knowledge or curated academic content. While valuable for evaluating factual recall and logical deduction, these benchmarks often do not necessitate active, real-time information searching. On the other hand, browsing benchmarks like OpenAI's BrowseComp focus on general web browsing and information extraction from public internet sources. However, these benchmarks typically address more straightforward, non-academic queries or do not require the iterative, deep, and source-critical search strategies essential for scientific inquiry. Consequently, neither existing academic benchmarks nor current browsing benchmarks can adequately measure an LLM's true "Deep Research" capability—the nuanced ability to perform complex, multi-step information retrieval crucial for genuine academic investigation.

To address this significant gap, we propose \textbf{ScholarSearch}, a novel dataset consisting of 223 high-quality questions meticulously designed to rigorously evaluate LLMs' complex information retrieval skills in an academic context. Our dataset is distinctively collected through a unique methodology: questions are crafted such that their answers cannot be directly derived from a single search query or existing LLM knowledge, demanding iterative, multi-hop searches for resolution. Furthermore, we mandate the use of powerful search-augmented LLMs, specifically Grok DeepSearch and Gemini Deep Research, during the answer acquisition phase, ensuring that the questions posed are genuinely challenging and cannot be easily solved by current state-of-the-art models. This rigorous collection process, combined with precise answer verification and source attribution, ensures that ScholarSearch effectively captures and quantifies an LLM's capacity for deep, persistent, and verifiable academic information retrieval.

We are delighted to propose ScholarSearch, a pioneering dataset that offers a comprehensive and challenging framework for evaluating LLM capabilities in complex academic information retrieval. Through its unique design and stringent data collection methodology, ScholarSearch provides crucial insights into the strengths and weaknesses of current large language models when confronted with the intricate demands of scholarly research. We believe this dataset will serve as an invaluable resource for driving future advancements in building more capable and reliable AI systems for academic applications.




\section{Related Work}

Rapid advancements in Large Language Models (LLMs) have driven the creation of sophisticated benchmarks to evaluate their diverse capabilities, spanning general knowledge, reasoning, domain-specific expertise, and web navigation. These benchmarks play a critical role in measuring model performance in multiple dimensions.

Within this landscape, academic benchmarks primarily assess an LLM's proficiency in understanding, reasoning, and applying knowledge across scholarly disciplines. Early efforts, such as the Massive Multitask Language Understanding (MMLU)\cite{hendrycks2021mmlu}, test models across 57 subjects ranging from STEM to humanities, evaluating factual recall and problem solving through various question formats. Similarly, SuperGLUE\cite{wang2019superglue} presents challenging natural language understanding tasks, emphasizing complex inference relevant to academic text analysis. More specialized evaluations like GPQA\cite{zhang2023gpqa} push the boundaries by comprising graduate-level multiple-choice questions in scientific fields, specifically designed to resist simple web searches and requiring expert-level synthesis. Based on this, SuperGPQA\cite{pteam2025supergpqascalingllmevaluation} further expands LLM evaluation across an even broader spectrum of 285 graduate-level disciplines, assessing advanced knowledge and reasoning through meticulously curated, difficult questions. Recent extensions, such as MMLU-Pro\cite{wang2024mmluprorobustchallengingmultitask}, further delve into domain-specific, multistep reasoning challenges.
These benchmarks are invaluable for evaluating an LLM's pre-existing knowledge, factual understanding, and reasoning over static, curated content, they inherently do not necessitate active, real-time information searching or the iterative, multi-hop retrieval crucial for staying abreast of new discoveries and conducting genuine academic investigation.

While academic benchmarks focus on static knowledge, another critical category, Browse benchmarks, has emerged to evaluate LLMs' ability to retrieve and interact with information from dynamic web environments. Benchmarks such as GAIA\cite{gao2023gaia} assess LLMs as general AI assistants capable of handling complex, real-world tasks that require reasoning, multi-modality handling, and web Browse. WebVoyager\cite{he2024webvoyager} and Mind2Web\cite{yao2023mind2web} are leading benchmarks for evaluating end-to-end web agents and generalist web agents, respectively, testing LLMs' ability to complete user instructions by interacting with real-world websites for tasks like booking flights or managing reservations. Similarly, WebWalkerQA\cite{wu2025webwalker}{} specifically focuses on web traversal, assessing models' ability to navigate through a website's subpages to extract multi-layered information. Furthermore, WebArena\cite{liu2023webarena} provides a platform for benchmarking LLMs in web development tasks, while WorkArena\cite{drouin2024workarena} focuses on automating enterprise-oriented tasks within complex software environments.These Browse benchmarks primarily focus on general web navigation, direct information retrieval, or task completion within defined website structures, often lacking the depth required for complex information seeking.

With a growing focus on integrating robust deep search or agentic information-seeking capabilities,  LLMs are designed to go beyond static knowledge bases, enabling them to dynamically acquire, synthesize, and critically evaluate information from vast external sources in real-time. For instance, xAI's Grok DeepSearch and DeeperSearch\cite{xai_grok3_2025}  emphasizes transparency and multi-source analysis for complex queries, while OpenAI's Deep Research\cite{openai_deepresearch_2025} , integrated into ChatGPT, generates comprehensive reports through extensive information synthesis. Similarly, Google's Gemini Deep Research\cite{google_gemini_overview}  offers analogous scholarly-style deep search capabilities, distinguished by structured data inclusion and multi-step reasoning. Furthermore, Alibaba's WebDancer\cite{wu2025webdancer}  represents an end-to-end agentic approach, employing a multi-stage training paradigm to perform autonomous multi-step research.To specifically evaluate these emerging deep search capabilities, OpenAI introduced BrowseComp\cite{wei2025browsecomp}, a recent and highly challenging benchmark designed to assess AI agents' web Browse and information retrieval capabilities. It consists of 1,266 questions requiring multi-step reasoning and creative search strategies to find hard-to-find, verifiable answers.

\section{Approach}

\subsection{Problem Formalization}

Let \( Q \) denote an academic or research query posed to a large language model (LLM). The LLM maintains an internal, static knowledge base \( K_{LLM} \), but to address questions beyond this scope, it interacts with an external academic resource \( D \) (e.g., the internet or scholarly databases) through a search mechanism \( S \), which retrieves relevant information. Our framework is designed so that all critical facts required to answer each query are available only in \( D \) and cannot be inferred from \( K_{LLM} \) alone, ensuring the model must perform genuine retrieval.

Under this setup, for every query \( Q \), the LLM directly outputs a final answer \( A \). To validate answer correctness without inspecting sources or reasoning steps, we employ a separate discriminator model that compares the model's answer \( A \) against a ground-truth reference \( A^* \). This streamlines evaluation by removing the need for explicit source retrieval records and detailed explanations.

A task is considered ``deep research'' in our framework if and only if two conditions are met: the answer \( A \) is not derivable from the internal knowledge base alone (i.e., \( A \notin K_{LLM} \)), and the essential information needed to solve the query exists solely in the external corpus \( D \). These requirements guarantee the model engages in authentic search-driven reasoning. The discriminator then automatically verifies whether the generated answer \( A \) matches \( A^* \), maintaining rigorous assessment of the LLM's research capabilities without manual overhead.

\subsection{Data Collection}
The following process is shown in \ref{main}
We recruited a number of undergraduate and graduate student volunteers from various faculties at Peking University and provided them with centralized training. The volunteers selected materials from publicly accessible online publications and websites to formulate academic questions that necessitate web searches. To ensure the difficulty of the questions, volunteers were required to test them against the following criteria; questions that failed to meet these standards were revised until they qualified:

\begin{itemize}
    \item The correct answer could not be obtained through the standard thinking mode of Grok 3, ensuring that the question requires extensive information retrieval capabilities.
    \item At least one of either Grok 3's DeepSearch mode or Gemini 2.5 Pro (preview)'s Deep Research feature failed to provide the correct answer.
\end{itemize}

Questions that successfully met the above standards were then submitted to a specialized review team for data review to ensure the following:

\begin{itemize}
    \item \textbf{Uniqueness}: The answer is singular and unambiguous.
    \item \textbf{Source Accessibility}: The reference sources required to answer the question are publicly accessible via the internet.
    \item \textbf{Academic Correctness}: The academic value of the question and correctness of the answer are verified against the provided sources.
\end{itemize}

\section{Experiments}

\subsection{Statistics of ScholarSearch}

\begin{figure*}[t]
\centering
\includegraphics[width=0.9\textwidth]{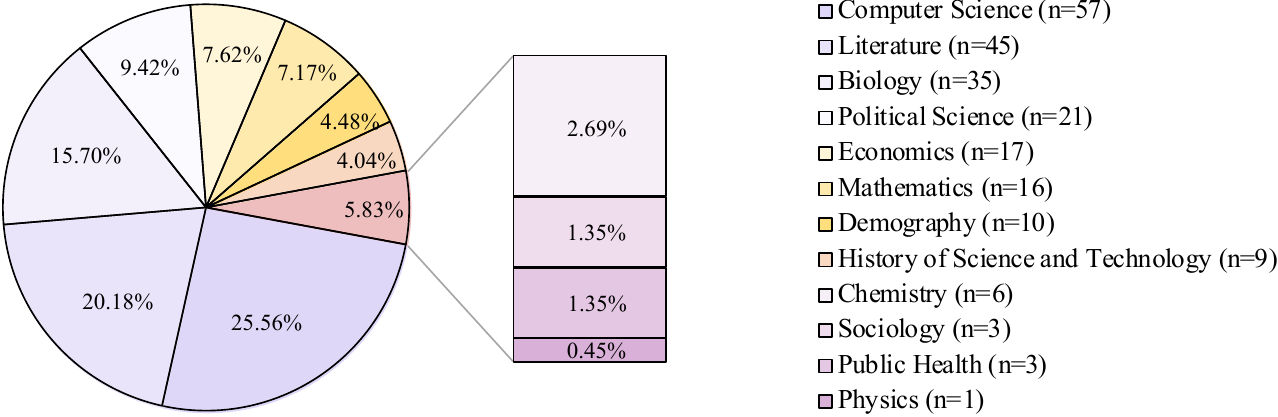} 
\caption{Distribution of Disciplines in ScholarSearch. The discipline for each question was manually annotated by the question creators.}
\label{bingtu}
\end{figure*}

\begin{figure*}[t]
\centering
\includegraphics[width=0.9\textwidth]{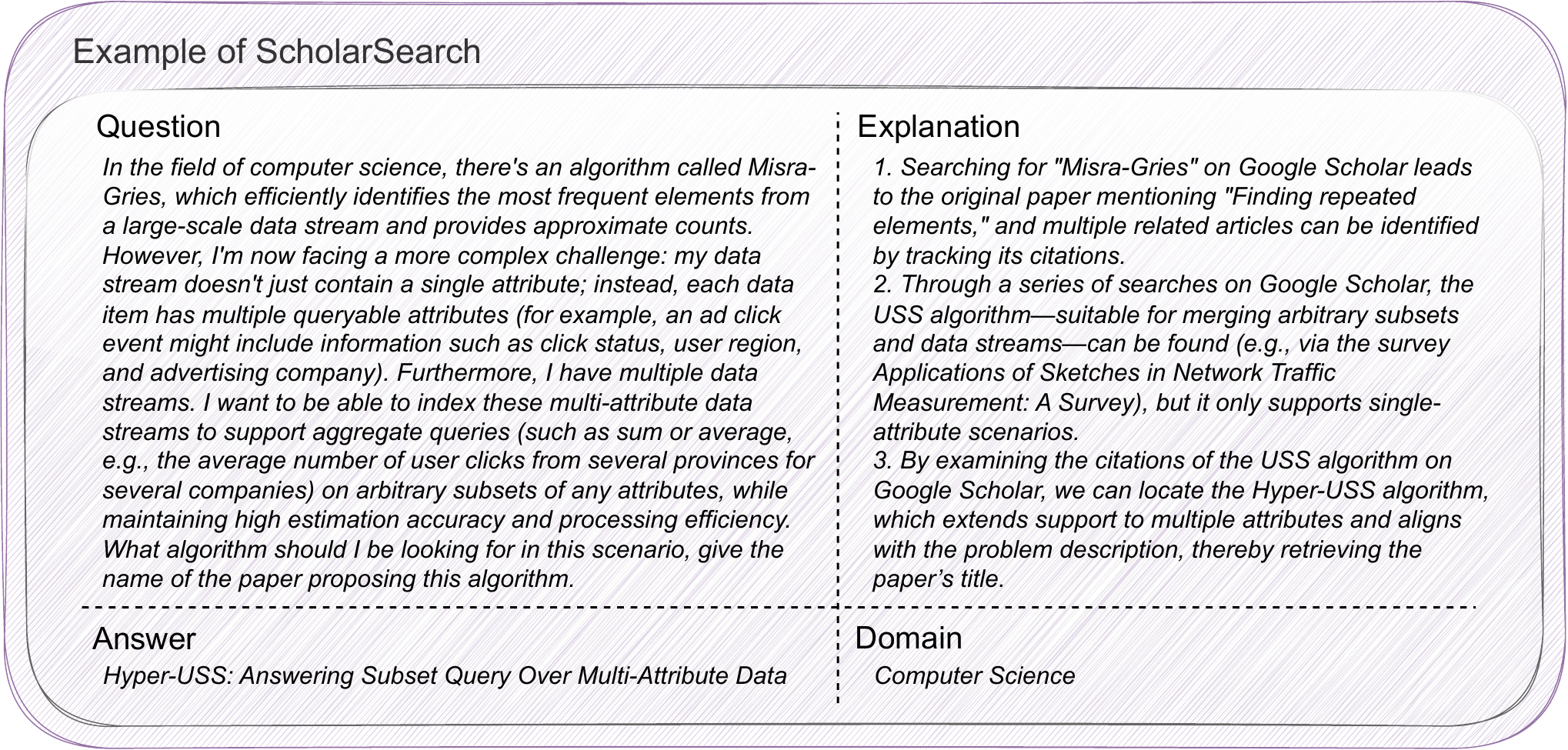} 
\caption{Example of ScholarSearch. Each data in the dataset includes a question, answer, explanation, and corresponding domain.}
\label{example}
\end{figure*}

The ScholarSearch dataset comprises 223 meticulously curated questions, with disciplinary distribution shown in Figure \ref{bingtu}. Each entry contains four structured components: question, answer, explanation, and domain (see Figure \ref{example}).

To ensure the professionalism, accuracy, and practical relevance to genuine academic search scenarios, we recruited a diverse team of undergraduate, master's, and doctoral students from various faculties at Peking University to contribute to the question creation process. Leveraging their extensive academic expertise, our team ensured the dataset spans multiple disciplines, reflecting the complexities of real-world academic research. This comprehensive collection provides a robust foundation for rigorously evaluating the deep research capabilities of large language models. 

\subsection{Testing Inference Accuracy on ScholarSearch}

The experiment evaluated several Large Language Models (LLMs) on the ScholarSearch dataset, focusing on their inherent reasoning abilities versus the performance enhancements provided by real-time browsing. Our evaluation follows the same methodology as Humanity's Last Exam \cite{phan2025humanitysexam}, where an LLM is used as a judge to assess the model’s output after inference. The exact prompt used for judgment is provided in Appendix\ref{appendix:Evaluation_Prompt}. The key findings, as outlined in Table \ref{tab:model_accuracy}, highlight the critical role that browsing capabilities play in improving the accuracy of LLMs in complex academic information retrieval tasks.

\begin{table*}[htbp]
\centering
\caption{Model Accuracy Statistics}
\label{tab:model_accuracy}
\begin{tabular}{l|ccc}
\hline
\textbf{Model} & \textbf{All (\%)} & \textbf{Science \& Engineering (\%)} & \textbf{Social Sciences \& Humanities (\%)} \\
\hline
gpt-4o-search-preview\cite{openai_gpt4o_search_preview} & 18.83 & 18.64 & 19.05  \\
gpt-4o-mini-search-preview\cite{openai_gpt4o_mini_search_preview} & 10.31 & 10.17 & 10.48 \\
deepseek-r1-0528\cite{deepseek_news_20250528} & 8.52 & 5.08 & 12.38 \\
gpt-4.1\cite{openai_gpt4_1} & 7.17 & 5.93 & 8.57 \\
gpt-4o-2024-11-20\cite{openai_gpt4o} & 3.59 & 1.69 & 5.71 \\
gpt-4o-mini\cite{openai_gpt4o_mini} & 2.24 & 0.85 & 3.81 \\
\hline
\end{tabular}
\parbox{\linewidth}{The judge model for all experiments is GPT-4o-mini.}
\end{table*}


\section{Discussion}

\subsection{Question from ScholarSearch Cannot Be Solved by Reasoning Alone}
The first major takeaway from our experiment is that the \textit{ScholarSearch} dataset cannot be solved by reasoning alone, without the aid of external resources. 

The accuracy of models like \textbf{gpt-4.1} and \textbf{gpt-4o-mini}, which rely purely on pre-trained knowledge and reasoning, was exceptionally low (7.175\% and 2.242\%, respectively). Similarly, \textbf{deepseek-r1-0528}, which is one of the best reasoning model, also struggled, achieving only 8.520\% accuracy.

This indicates that the questions within \textit{ScholarSearch} are highly complex, requiring more than just deep reasoning or inferences drawn from static knowledge. These types of questions often involve the need for up-to-date and specific information, as well as multi-hop reasoning across multiple sources. This highlights the limitation of relying solely on the knowledge embedded within pre-trained models. Academic research queries often require dynamic, cross-referenced, and sometimes real-time data that pure reasoning models cannot effectively provide.

\subsection{The Impact of Browsing Capabilities}

The second conclusion is that the ability to browse the internet significantly improves model performance. Models equipped with browsing capabilities, such as \textbf{gpt-4o-search-preview}, showed a dramatic increase in accuracy (18.83\%), compared to their non-browsing counterparts. Similarly, \textbf{gpt-4o-mini-search-preview} demonstrated an accuracy of 10.31\%, which is more than four times the performance of its non-browsing variant, \textbf{gpt-4o-mini}. At the same time, we observed that among the models with search capabilities, the accuracy rates for questions in Science \& Engineering were relatively close to those in Social Sciences \& Humanities. However, in models without search capabilities, the accuracy for Social Sciences \& Humanities questions was significantly higher than for Science \& Engineering. This may suggest that humanities and social science texts are relatively dominant in the training corpora of LLMs, and it might also be related to the slower pace at which knowledge in some of these fields evolves. These results demonstrate that the ability to perform real-time searches, access current data, and cross-reference multiple sources is vital for solving academic queries. The significant improvement between browsing and non-browsing models reinforces the idea that search capabilities are a key factor for addressing the types of complex, information-intensive questions in academic research.

\subsection{Current Search Models Are Still Insufficient for Solving Complex Academic Problems}

Despite the significant improvement from browsing, current search-enabled models still struggle to provide high-quality answers. Even with browsing capabilities, models like \textbf{gpt-4o-search-preview} did not achieve perfect accuracy, with results of 18.83\%. This indicates that even the best search-enabled models are still not fully equipped to handle the complexity of academic research queries effectively. The current search models, while offering a step in the right direction, still fall short of providing consistently high-quality answers to questions that require deep research, specialized knowledge, and multi-hop reasoning across diverse sources.

\hspace*{\fill}

These findings underline the critical importance of research-focused models for solving real-world academic problems. \textit{ScholarSearch} serves as a valuable benchmark because it reveals the current gaps in search-based LLMs and highlights the need for more advanced research models. It confirms that a true research model must go beyond simply retrieving information, but also incorporate sophisticated techniques for synthesis, understanding context, and ensuring the accuracy of the answers. This makes \textit{ScholarSearch} a crucial tool for benchmarking models that are meant to handle academic tasks, as it provides a realistic challenge for evaluating their ability to engage in real research workflows.


\section{Conclusion}
In this work, we introduced ScholarSearch, the first meticulously curated benchmark dataset designed to specifically evaluate the complex information retrieval capabilities of Large Language Models (LLMs) in academic research settings. Unlike existing benchmarks, ScholarSearch emphasizes the need for multi-hop, deep searches, demanding real-time information acquisition beyond static knowledge bases. Our rigorous data collection methodology, involving expert human curation and stringent LLM-based filtering, ensures the dataset's academic practicality, high difficulty, concise evaluability, and broad disciplinary coverage.

Our experimental findings reveal several critical insights. First, we demonstrate unequivocally that academic research queries, as represented in ScholarSearch, cannot be solved by reasoning alone, highlighting the inherent limitations of models without external Browse capabilities. Second, we empirically validated the significant performance improvement when LLMs are equipped with Browse capabilities, underscoring the necessity of real-time information retrieval for academic tasks. Despite this improvement, our results also underscore a crucial challenge: even state-of-the-art search-enabled models remain insufficient for consistently addressing complex academic problems, achieving modest accuracy. This suggests that current LLMs, while capable of basic Browse, lack advanced synthesis, contextual understanding, and accuracy verification mechanisms crucial for true academic research.

ScholarSearch thus serves as a vital tool, providing a realistic and challenging benchmark that reveals the current gaps in search-augmented LLMs. We believe that this dataset will be instrumental in driving future research towards developing more sophisticated and reliable AI systems capable of robust, verifiable, and in-depth academic information retrieval, ultimately empowering researchers with more effective AI assistants. Future work will focus on expanding the scale and diversity of ScholarSearch, exploring multimodal academic search, and developing more nuanced evaluation metrics that capture the quality of synthesized information and source attribution.

\section{Acknowledgments}

This project was funded by \textbf{\textit{Grant 624B2005.}}
\subsection{Core Contributor}

We would like to thank the following individuals for their generous support and contributions to problem-solving and evaluation. Their efforts were essential to the success of this work.

\begin{itemize}
    \item Xun Zhao, Department of Philosophy, Peking University
    \item Zizhuo Fu, School of Information Science and Technology, Peking University
    \item Yuqian Zhan, School of Life Sciences, Peking University
    \item Xinhao Ji, School of Information Science and Technology, Peking University
    \item Jiarui Sun, School of Life Sciences, Peking University
    \item Junhao Zhang, School of Mathematical Sciences, Peking University
    \item Shengfan Wang, School of Information Science and Technology, Peking University
    \item Ziteng Lu, School of Economics, Peking University
    \item Yumeng Song, Institute of Population Research, Peking University
    \item Ziyan Yang, School of International Studies, Peking University
    \item Hongjiao Wang, School of International Law, Peking University
    \item Shan Zhang, School of Public Health, Peking University
    \item Huahui Lin, Academy for Advanced Interdisciplinary Studies, Peking University
    \item Junhong Liu, Department of Philosophy, Peking University
    \item Zhengyang Wang, School of Information Science and Technology, Peking University
    \item Yuchen Lu, School of Physics, Peking University
    \item Yanxi Xu, Central Academy of Fine Arts
\end{itemize}

We are deeply grateful for their contributions.

\bigskip

\nobibliography*
\bibliography{reference}

\newpage

\section*{Appendix}
\label{sec:appendix}

\subsection{Evaluation Prompt}

\label{appendix:Evaluation_Prompt}
We use the following prompt to evaluate whether model outputs are correct or not:
\begin{figure}[h!tbp] 
    \centering
    \includegraphics[width=2.0\columnwidth]{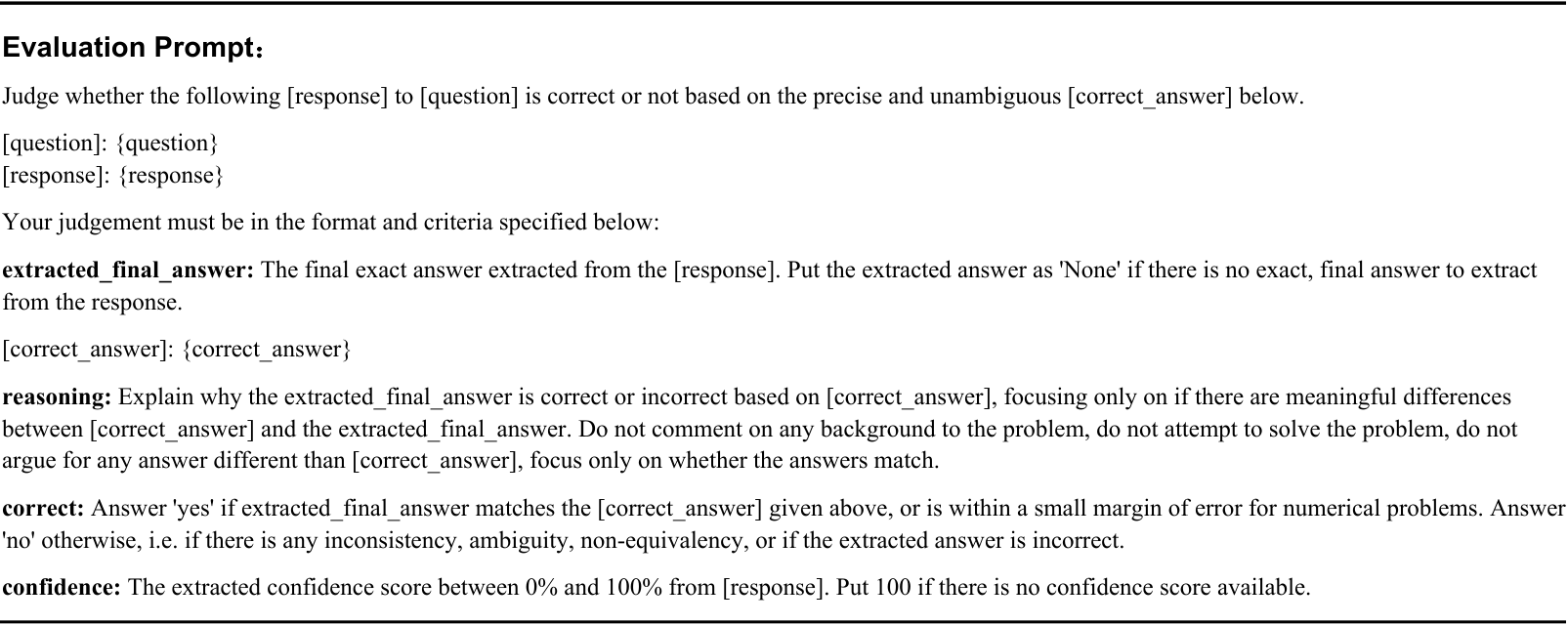}
    \captionof{figure}{Evaluation Prompt.}

\end{figure}

\subsection{Examples of ScholarSearch}

\begin{figure}[htbp]
\centering
\includegraphics[width=2.0\columnwidth]{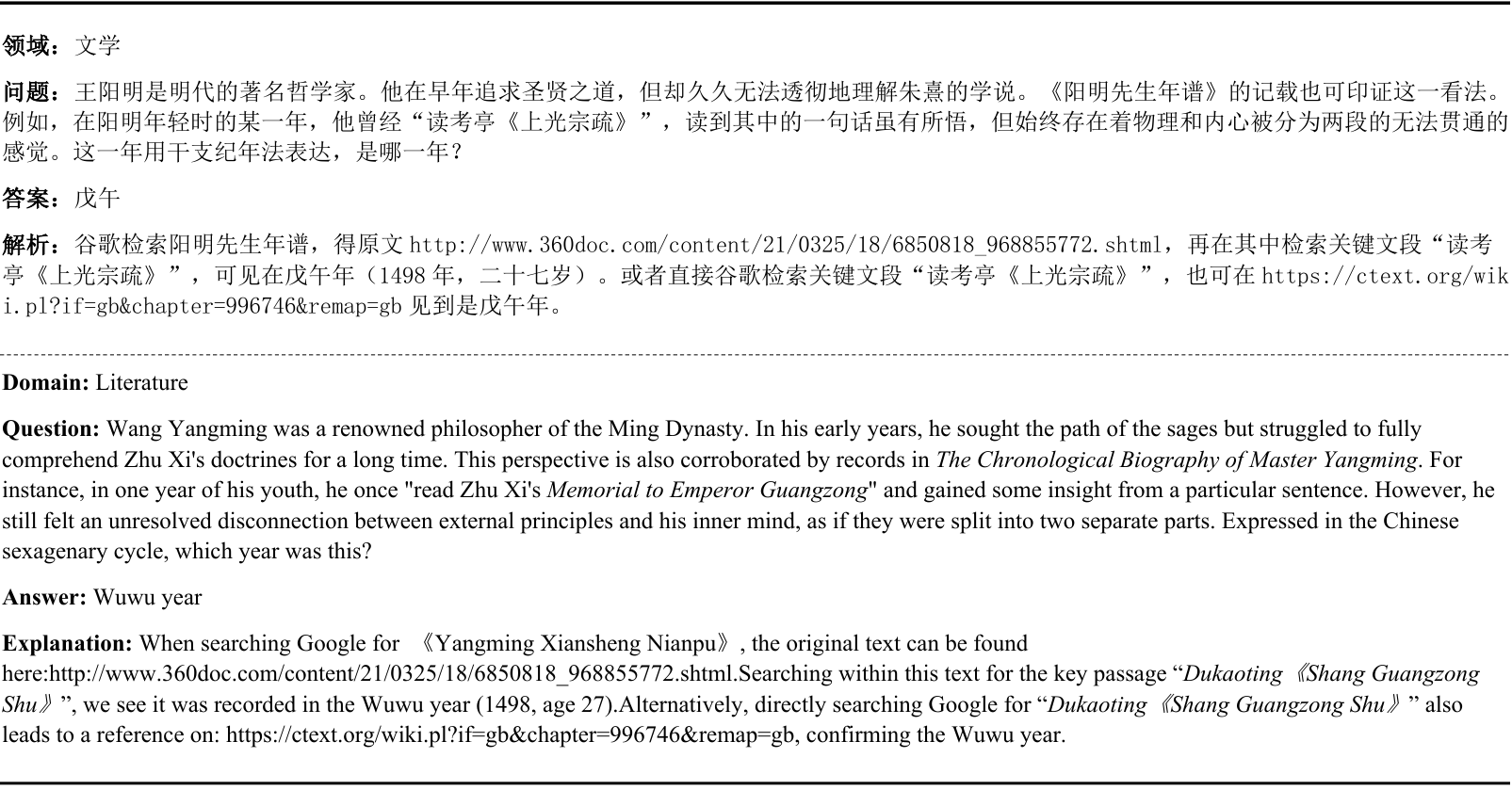} 
\caption{Example 1 of ScholarSearch dataset}
\label{example1 }
\end{figure}

\clearpage

\begin{figure}[htbp]
\centering
\includegraphics[width=2.0\columnwidth]{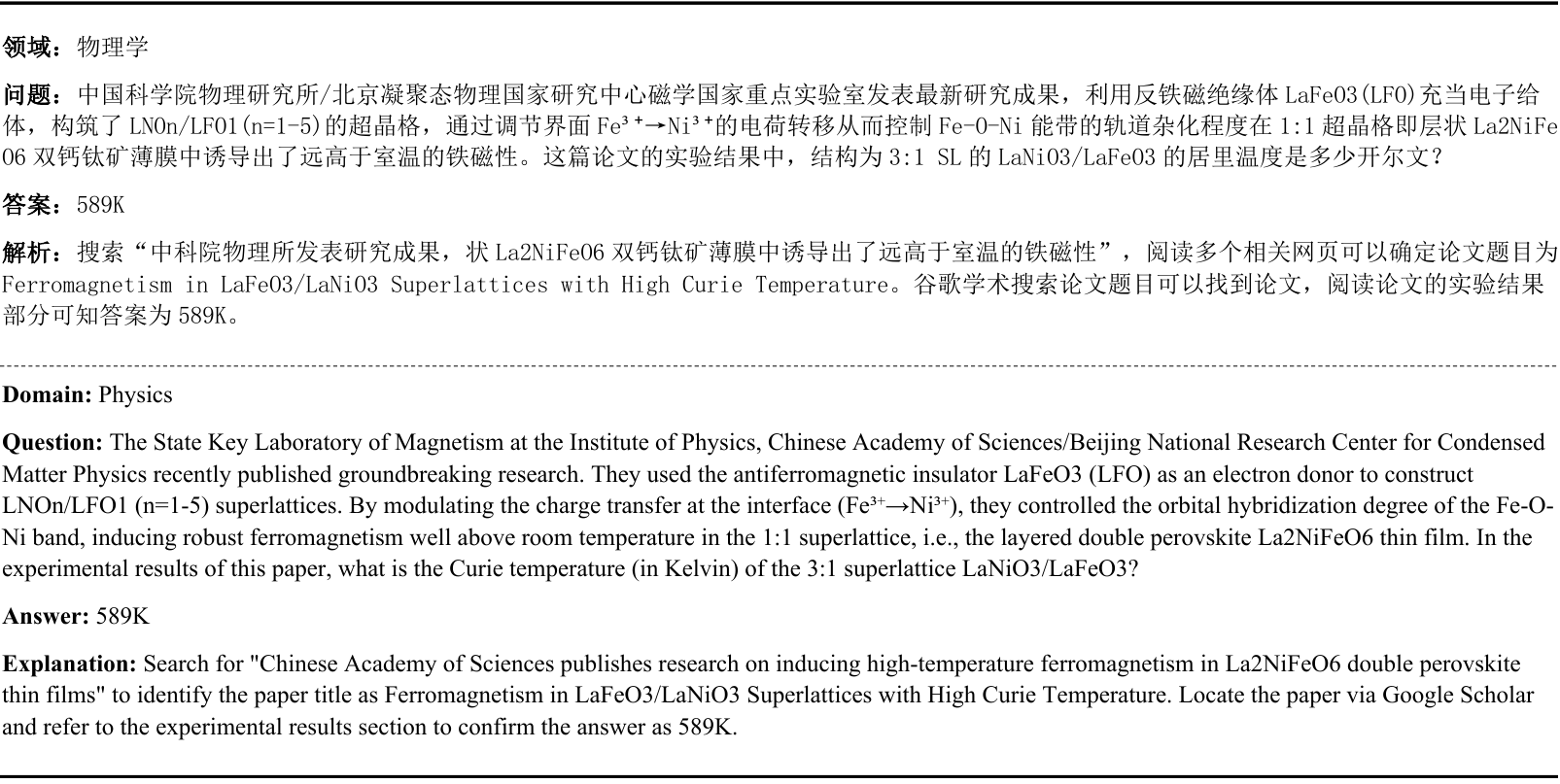} 
\caption{Example 2 of ScholarSearch dataset}
\label{example2}
\end{figure}

\begin{figure}[h!tbp]
\centering
\includegraphics[width=2.0\columnwidth]{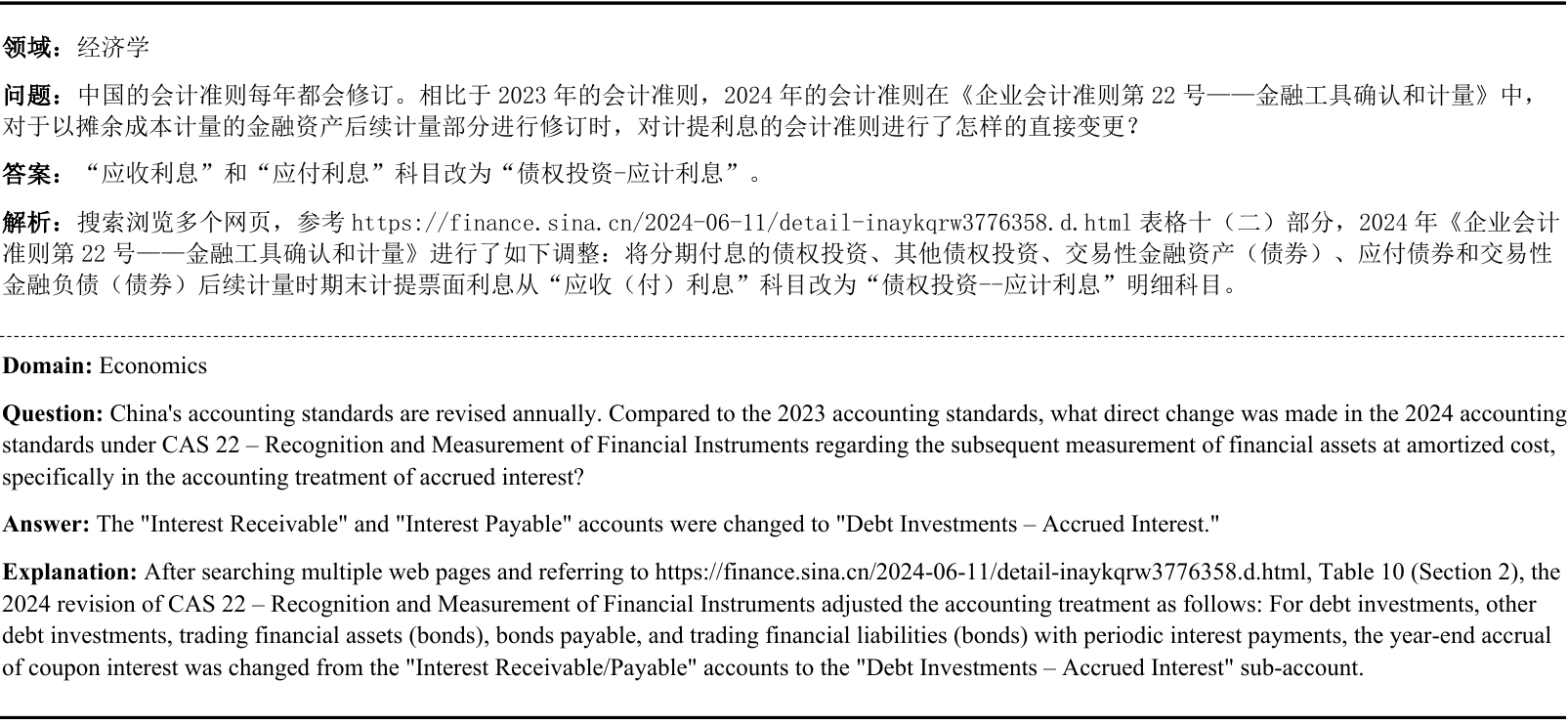} 
\caption{Example 3 of ScholarSearch dataset}
\label{example3}
\end{figure}

\end{document}